%% file: paper.tex
\documentclass{article}
\usepackage{times}
\usepackage{subfigure}
\usepackage{color}
\usepackage{graphicx,amssymb,amsmath}
\usepackage{xspace}
\usepackage[ruled, vlined]{algorithm2e} % [section] is use to define the numbering mode

\newcommand{\commentout}[1]{}
\newcommand{\eat}[1]{}
\newcommand{\topic}[1]{\vspace{10pt}\noindent{\underline{#1} :}}

\newcommand{\calC}{{\mathcal C}}
\newcommand{\calF}{{\mathcal F}}
\newcommand{\calP}{{\mathcal P}}
\newcommand{\calN}{{\mathcal N}}

\newcommand{\dist}{\mathsf{dist}}
\newcommand{\depth}{\mathsf{depth}}

\newcommand{\B}{\mathsf{B}}

\newcommand{\lMCenter}{$\ell$-D{\footnotesize IVERSITY}}
\newcommand{\lMCenterOutlier}{$\ell$-D{\footnotesize
    IVERSITY}-O{\footnotesize UTLIERS}}

\newcommand{\rgather}{$r$-G{\footnotesize ATHERING}}

\newcommand{\ldiversity}{$\ell$-D{\footnotesize IVERSITY}}
\newcommand{\kanonymity}{$k$-A{\footnotesize NONYMITY}}

\newcommand{\flowtest}{F-T{\footnotesize EST}}
\newcommand{\matchtest}{M-T{\footnotesize EST}}

\newenvironment{proof}{\noindent {\em Proof: }\ignorespaces}{}
\newcommand{\qed}{\hspace*{\fill}$\Box$\medskip}

\newtheorem{theorem}{Theorem}
\newtheorem{lemma}{Lemma}

\newtheorem{definition}{Definition}
\newtheorem{property}{Property}

\title{Clustering with Diversity}

\author{Jian Li$^1$ \and Ke Yi$^2$ \and Qin Zhang$^2$}

\date{$^1$University of Maryland, College Park \\
{\tt lijian@cs.umd.edu} \\
$^2$Hong Kong University of Science and Technology \\
{\tt $\{$yike,qinzhang$\}$@cse.ust.hk}}

\begin{document}
\maketitle
\vspace{-0.7cm}
\begin{abstract}
  We consider the {\em clustering with diversity} problem: given a set of
  colored points in a metric space, partition them into clusters such that
  each cluster has at least $\ell$ points, all of which have distinct
  colors. 
  We give a 2-approximation to this problem for any $\ell$ when the
  objective is to minimize the maximum radius of any cluster.  We show
  that the approximation ratio is optimal unless
  $\mathbf{P=NP}$, by providing a matching lower bound.  
  Several extensions to our algorithm have also been
  developed for handling outliers.  This problem is
  mainly motivated by applications in privacy-preserving data publication. \\[3pt]
\end{abstract}

%\newpage

\input{intro}

%\input{preliminary}
\vspace{-0.3cm}
\input{algorithm}

\input{lowerBound}

\input{outlier}
\vspace{-0.3cm}
\bibliographystyle{abbrv}
\bibliography{paper}

%\newpage
%\appendix
%\input{appendix}

\end{document}

%% file: intro.tex
\section{Introduction}
\label{sec:intro}
Clustering is a fundamental problem with a long history and a rick collection of results. A general
clustering problem can be formulated as follows. Given a set of points $P$
in a metric space, partition $P$ into a set of disjoint clusters such that
a certain objective function is minimized, subject to
some cluster-level and/or instance-level constraints.
Typically, cluster-level constraints impose restrictions
on the number of clusters or on the size of each cluster.
The former corresponds to the classical $k$-center, $k$-median, $k$-means
problems, while the latter has recently received much attention from various research communities
\cite{Ji04,AFK+06,hoppner2008csc}.
%mainly motivated by applications in privacy preservation.
On the other hand, instance-level constraints specify
whether particular items are similar or dissimilar, usually based on some background knowledge
\cite{Wagstaff00clusteringwith,journal/jacm/ailon08}.
In this paper, we impose a natural instance-level constraint on a clustering problem, that the
points are colored and all points partitioned into one cluster must have
distinct colors.  We call such a problem {\em clustering with diversity}.
Note that the traditional clustering problem is a special case of ours
where all points have unique colors.

As an illustrating example, consider the problem of choosing
locations for a number of factories in an area where different resources
are scattered. Each factory needs at least $\ell$ different resources
allocated to it and the resource in one location can be sent to only one factory.  
This problem corresponds to our clustering problem where each
kind of resource has a distinct color, and we have a lower bound
$\ell$ on the the cluster size.

\begin{figure}[t]
    \centerline{\includegraphics[width=\linewidth]{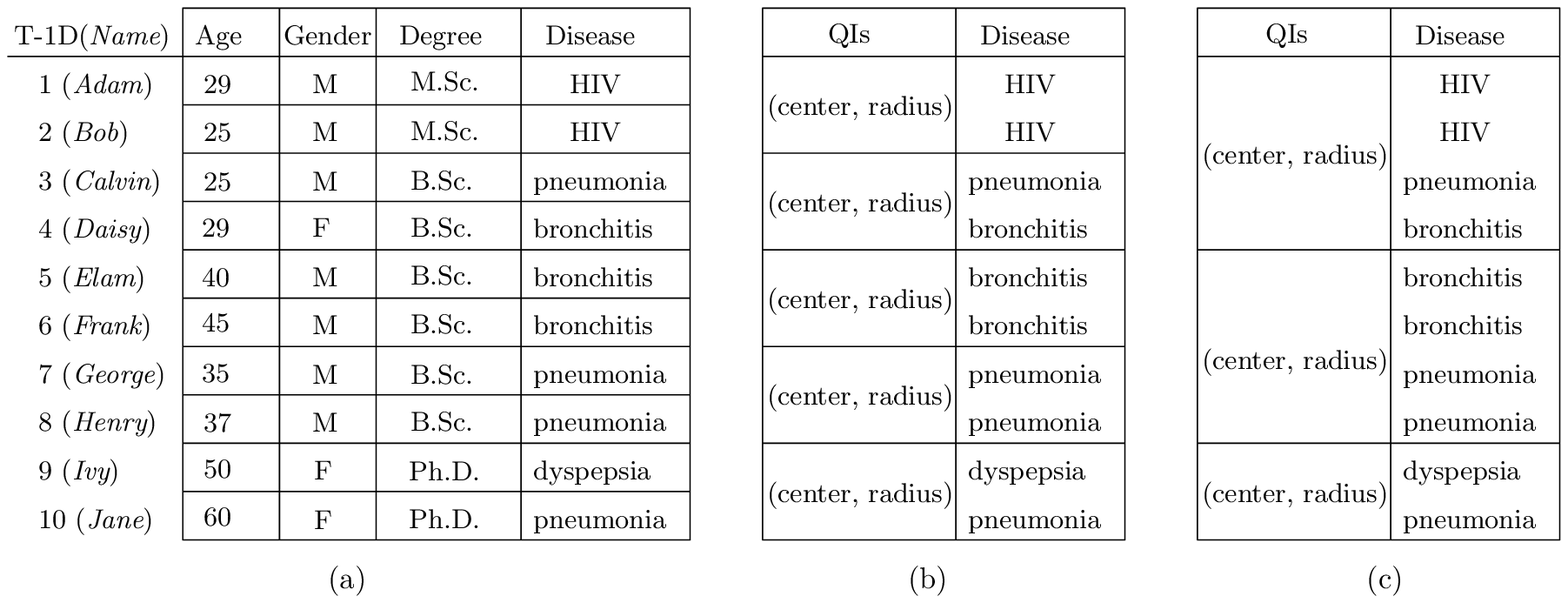}}
    \caption{(a) The microdata; (b) A $2$-anonymous table; (c) An $2$-diverse table;}
    \label{fig:microdata}
\end{figure}

The main motivation to study clustering with diversity is privacy
preservation for data publication, which has drawn tremendous attention in
recent years in both the database community
\cite{HK07,c.-w.wong06,xiao07,machanavajjhala06,xiao06:_anatom,beresford03,Xiao:10}
and the theory community
\cite{AFK+06,meyerson04,aggarwal05:_anony,feldman09:_privat,dwork09:_when}.
The goal of all the studies in privacy preservation is to prevent {\em
linking attacks} \cite{samarati01:_protec}.  Consider the table of
patient records in Figure~\ref{fig:microdata}(a), usually called the {\em
microdata}.  There are three types of attributes in a microdata table.
The {\em sensitive attribute} (SA), such as ``Disease'', is regarded as the
individuals' privacy, and is the target of protection.  The {\em
identifier}, in this case ``Name'', uniquely identifies a record, hence
must be ripped off before publishing the data.  The rest of the attributes,
such as ``Age'', ``Gender'', and ``Education'', should be published so that
researchers can apply data mining techniques to study the correlation
between these attributes and ``Disease''.  However, since these attributes
are public knowledge, they can often uniquely identify individuals
when combined together.  For example, if an attacker knows (i) the age
(25), gender (M), and education level (Master) of Bob, and (ii) Bob has a
record in the microdata, s/he easily finds out that Tuple 2 is Bob's record
and hence, Bob contracted HIV.  Therefore, these attributes are often
referred to as the {\em quasi-identifiers} (QI).  The solution is thus to
make these QIs ambiguous before publishing the data so that it is difficult
for an attacker to link an individual from the QIs to his/her SA, but at
the same time we want to minimize the amount of information loss due to the
ambiguity introduced to the QIs so that the interesting correlations
between the QIs and the SA are still preserved.

The usual approach taken to prevent linking attacks is to partition the
tuples into a number of {\em QI-groups}, namely clusters, and within each
cluster all the tuples share the same (ambiguous) QIs. There are various
ways to introduce ambiguity.  A popular approach, as taken by
\cite{AFK+06}, is to treat each tuple as a high-dimensional point in the
QI-space, and then only publish the center, the radius, and the number of
points of each cluster.  To ensure a certain level of privacy, each cluster
is required to have at least $k$ points so that the attacker is not able to
correctly identify an individual
with confidence larger than $1/k$.  This requirement is referred to
as the {\em \kanonymity} principle \cite{AFK+06,meyerson04}.
%~\footnote{This
%requirement is usually called {\em $k$-anonymity} in the database
%literature \cite{meyerson04}, where $k$ is the same as $r$ here; we
%choose to adopt the term ``\rgather'' from \cite{AFK+08}.}.
%This way the attacker will not be able to correctly identify an individual with
%probability larger than $1/r$.
The problem, translated to a clustering problem, can be phrased as follows:
Cluster a set of points in a metric space, such that each cluster has at least $r$ points.
When the objective is to minimize the maximum radius of all clusters,
the problem is called \rgather\, and a 2-approximation is known \cite{AFK+06}.
%The authors also give a constant-approximation when the objective function is the sum of all the
%radii.

%\begin{figure}[t]
%\centering
%  \subfigure[Microdata conforming to 2-anonymity.]{
%   \includegraphics[width=5.5cm]{intro2}
%    \label{fig:anony}
%  }
%\hspace{1cm}
%  \subfigure[Microdata conforming to 2-diversity.]{
% \includegraphics[width=5.5cm]{intro3}
%    \label{fig:div}
%  }
%    \caption{Microdata conforming to 2-anonymity and 2-diversity.}
%\end{figure}

However, the \kanonymity\ principle suffers from the {\em homogeneity} problem:
%as the ultimate goal of privacy preservation is not to merely protect the identity
%of a tuple, but the value of the SA.
%If a clustering only conforms to \rgather,
A cluster may have too many tuples with the same SA value.
For example, in Figure~\ref{fig:microdata}(b), all tuples in QI-group 1, 3, and 4
respectively have the the same disease. Thus, the attacker
can infer what disease all the people within a QI-group have contracted without
identifying any individual record.  The above
problem has led to the development of many SA-aware principles. Among them,
{\em \ldiversity} \cite{machanavajjhala06} is the most widely deployed
\cite{ghinita07:_fast,machanavajjhala06,kifer06:_injec,xiao07,xiao06:_anatom,c.-w.wong06},
due to its simplicity and good privacy guarantee.  The principle demands
that, in each cluster, at most $1/\ell$ of its tuples can have the same SA
value.  Figure~\ref{fig:microdata}(c) shows a 2-diverse version of the
microdata.  In an $\ell$-diverse table, an
attacker can figure out the real SA value of the individual with confidence
no more than $1/\ell$.
%The larger $\ell$ is, the better the privacy is
%protected.  On the other hand, larger values of $\ell$ also lead to larger
%radii of the clusters, hence increasing the ambiguity of the published
%data.
% An easy observation is that, if a cluster is $\ell$-diverse and has
% $\ge 2\ell$ points, then we can further decompose it into smaller
%$\ell$-diverse clusters
%.  Repeating this process all the clusters can be decomposed into
%until every cluster has $\ge \ell$ points, all with
%distinct SA values.  
Treating the SA values as colors, this problem then exactly corresponds to
the clustering with diversity problem defined at the beginning, where we
have a lower bound $\ell$ on the cluster size.

In contrast to the many theoretical results for \rgather\ and \kanonymity\
\cite{AFK+06,aggarwal05:_anony,meyerson04}, no approximation algorithm with performance guarantees is
known for \ldiversity,
% despite its importance and superiority to the
%former in terms of the level of privacy protection.  On the other hand,
even though many heuristic solutions have been proposed
\cite{lefevre06:_mondr,ghinita07:_fast,machanavajjhala06}. %without guarantees.

\paragraph{Clustering with instance-level constraints and other related work.}
Clustering with instance-level constraints is a developing area and
begins to find many interesting applications in various areas such as bioinformatics
\cite{bairoch1997spp}, machine learning \cite{Wagstaff00clusteringwith,wagsta2001ckm}, 
data cleaning \cite{arasu2009large}, etc.
Wagstaff and Cardie in their seminal work \cite{Wagstaff00clusteringwith} considered the following two types of
instance-level hard constraints: A {\em must-link (ML)} constraint dictates that two particular points must be clustered
together and a {\em cannot-link (CL)} constraint requires they must be separated.
Many heuristics and variants have been developed subsequently, e.g. \cite{wagsta2001ckm,xing:dml},
and some hardness results with respect to minimizing the number of clusters
were also obtained \cite{davidson2007iac}. However, to the best of our knowledge,
no approximation algorithm with performance guarantee is known for any version of the problem.
We note that an $\ell$-diverse clustering can be seen as a special case
where nodes with the same color must satisfy CL constraints.

As opposed to the hard constraints imposed on any clustering,
the correlation clustering problem \cite{bansal2004correlation} considers soft and possibly conflicting constraints
and aims at minimizing the violation of the given constraints.
An instance of this problem can be
represented by a complete graph with each edge labeled (+) or
(-) for each pair of vertices, indicating that two vertices should be in the same
or different clusters, respectively. The goal is to cluster the elements so as to minimize
the number of disagreements, i.e., (-) edges within clusters and (+) edges crossing clusters.
The best known approximations for various versions of the problem are due to Ailon et al.~\cite{journal/jacm/ailon08}.
If the number of clusters is stipulated to be a small constant $k$, there is a polynomial time 
approximation scheme \cite{giotis2006correlation}. In the Dedupalog project, Arasu et al.~\cite{arasu2009large}
considered correlation clustering together with instance-level hard constraints, with the aim of  
de-duplicating entity references .

Approximation algorithms for clustering with
outliers were first considered by Charikar et al. \cite{charikar2001afl}.
The best known approximation factor for \rgather\ with outliers is $4$ due
to Aggrawal et al. \cite{AFK+06}.

%\newpage

\paragraph{Our results.}
In this paper, we give the first approximation algorithms to the clustering
with diversity problem.
% where each cluster must have at least $\ell$
%points.  The objective function is the maximum radius of all clusters.
We formally define the problem as follows.

\begin{definition}[\lMCenter]
\label{def:lmcenter}
Given a set of $n$ points in a metric space where each of them has a color,
cluster them into a set $\calC$ of clusters, such that each cluster has at
least $\ell$ points, and all of its points have distinct colors.
The goal is to minimize the maximum radius of any cluster.
\end{definition}

Our first result (Section~\ref{sec:algorithm}) is a 2-approximation
algorithm for \lMCenter.  The algorithm follows a similar framework as in
\cite{AFK+06}, but it is substantially more complicated.  The difficulty is
mainly due to the requirement to resolve the conflicting colors in each cluster
while maintaining its minimum size $\ell$.
To the best of our knowledge, this is first approximation algorithm for
a clustering problem with instance-level hard constraints.
%The approximation ratio holds when we compare with the optimal clustering in
%which the cluster centers could be any points in the metric space.

Next, we show that this approximation ratio is the best possible by presenting a
matching lower bound (Section~\ref{sec:lowerbound}).
A lower bound of 2 is also given in \cite{AFK+06} for \rgather. But to
carry that result over to \lMCenter, all the points need to have
unique colors.  This severely limits to applicability of this hardness
result.  In Section~\ref{sec:lowerbound} we give a construction showing that even
with only 3 colors, the problem is NP-hard to approximate within any factor strictly less than $2$.
In fact, if there are only 2 colors, we show that the problem can be solved
optimally in polynomial time via bipartite matching.

Unlike \rgather, an instance to the \lMCenter\ problem may not have a
feasible solution at all, depending on the color distribution.  In
particular, we can easily see that no feasible clustering exists
when there is one color that has more than $\lfloor
n/\ell \rfloor$ points.
%no clustering exists that satisfies the conditions
%in Definition~\ref{def:lmcenter}.
%This, however, does not mean that the data is unpublishable.
One way to get around this problem is to have some points not
clustered (which corresponds to deleting a few records in the \ldiversity\ problem).
Deleting records causes information loss in the published data,
hence should be minimized. Ideally, we would like to delete points just
enough such that the remaining points admit a feasible $\ell$-diverse
clustering.  In Section~\ref{sec:outliers}, we consider the {\em
\lMCenterOutlier} problem, where we
%to minimize the maximum radius of the
compute an $\ell$-diverse clustering after removing the
least possible number of points. We give an $O(1)$-approximation algorithm
to this problem.

Our techniques for dealing with diversity and cluster size constraints may
be useful in developing approximation algorithms
for clustering with more general instance-level constraints.
%Due to space constraints, we only provide complete details for our first result.
%We refer interested readers to the full version of the paper for all missing details and proofs \cite{Li10diversityfull}.

%Note that \lMCenter\ is a special case of this problem,
%where we do not need to remove any points.

%% It is also possible to consider the {\sc $\ell$-Diversity-Median} problem,
%% where the objective function is the sum of the radii of the clusters.
%% However, this seems to be much more difficult, and we leave it as an
%% interesting open problem.

%%% Local Variables:
%%% mode: latex
%%% TeX-master: "paper"
%%% End:

%% file: algorithm.tex
\section{A 2-Approximation for \lMCenter}
\label{sec:algorithm}

%\red{we need the definition of $r$-gathering somewhere in the introduction or problem statement.}
In this section we assume that a feasible solution on a given input always
exists.  We first introduce a few notations.  Given a set of $n$ points in
a metric space, we construct a weighted graph $G(V,E)$ where $V$ is the set
of points and each vertex $v \in V$ has a color $c(v)$.  For each pair of
vertices $u,v \in V$ with different colors, we have an edge $e=(u,v)$, and
its weight $w(e)$ is just their distance in the metric space. For any $u,v
\in V$, let $\mathtt{dist}_G(u,v)$ be the shortest path distance of $u,v$
in graph $G$. For any set $A\subseteq V$, let $N_G(A)$ be the set of
neighbors of $A$ in $G$.  For a pair of sets $A\subseteq V, B\subseteq V$,
let $E_G(A;B) = \{(a,b)\ |\ a\in A, b\in B, (a,b) \in E(G)\}$. The diameter
of a cluster $C$ of nodes is defined to be $d(C)=\max_{u,v\in C}(w(e(u,v)))$. 
Given a cluster $C$ and its center $v$, the
radius $r(C)$ of $C$ is defined as maximum distance from any node of $C$ to $v$,
i.e., $r(C)=\max_{u\in C}w(u,v)$. By triangle inequality, it is obvious 
to see that $\frac{1}{2}d(C)\leq r(C)\leq d(C)$.

A {\em star forest} is a forest where each connected component is a star. A
{\em spanning star forest} is a star forest spanning all vertices.  The
{\em cost} of a spanning forest $\calF$ is the length of the longest edge
in $\calF$. We call a star forest {\em semi-valid} if each star component
contains at least $\ell$ colors and {\em valid} if it is semi-valid and
each star is {\em polychromatic}, i.e., each node in the star has a
distinct color.
Note that a spanning star forest with cost $R$ naturally defines a
clustering with largest radius $R$.
%% Therefore, the classic \kcenter\ problem
%% can be rephrased as finding a min-cost spanning forest with at most $k$
%% stars.  The \lMCenter\ problem is to find a min-cost valid spanning forest.
Denote the radius and the diameter of the optimal clustering by $r^{*}$ and $d^{*}$, respectively.

We first briefly review the 2-approximation algorithm for the
\rgather\ problem \cite{AFK+06},
which is the special case of our problem when all
the points have distinct colors.  Let $e_1, e_2, \dots$ be the
edges of $G$ in a non-decreasing order of their weights.  The
general idea of the \rgather\ algorithm \cite{AFK+06} is to
first guess the optimal radius $R$ by considering each graph $G_i$
formed by the first $i$ edges $E_i=\{e_1,\dots,e_i\}$, as
$i=1,2,\dots$. It is easy to see that the cost of a spanning star
forest of $G_i$ is at most $w(e_i)$. For each $G_i\ (1 \le i \le
m)$, the following condition is tested (rephrased to fit into our
context):
\begin{enumerate}
\item[(I)] There exists a maximal independent set $I$ such that there is a
  spanning star forest in $G_i$ with the nodes in $I$ being the star
  centers, and each star has at least $r$ nodes.
\end{enumerate}
It is proved \cite{AFK+06} that the condition is met if the length of $e_i$
is $d^*$. The condition implies the radius of our solution is at most $d^{*}$
which is at most $2r^{*}$. 
Therefore, we get an 2-approximation.  
In fact, the independent set $I$ can be chosen greedily
and finding the spanning star forest can be done via a network
flow computation.

Our 2-approximation for the $\ell$-diversity problem follows the same
framework, that is, we check each $G_i$ in order and test the following
condition:
\vspace{-0.1cm}
\begin{enumerate}
\item[(II)] There exists a maximal independent set $I$ such that there is a
  {\em valid} spanning star forest in $G_i$ with the nodes in $I$ being the
  star centers.
\end{enumerate}

The additional challenge is of course that, while condition (I) only puts a
constraint on the size of each star, condition (II) requires both the size
of each star to be at least $\ell$ and all the nodes in a star have
distinct colors.  Below we first give a constructive algorithm that for a
given $G_i$, tries to find an $I$ such that condition (II) is met.  Next we
show that when $w(e_i) = d^*$, the algorithm is guaranteed to succeed.
The approximation ratio of 2 then follows immediately.

To find an $I$ to meet condition (II), the algorithm starts with an
arbitrary maximal independent set $I$, and iteratively augments it until
the condition is met, or fails otherwise.  In each iteration, we maintain
two tests.  The first one, denoted by {\em flow test} F-T{\footnotesize
  EST}$(G_i,I)$, checks if there exists a semi-valid spanning star forest
in $G_i$ with nodes in $I$ being star centers.  If $I$ does not pass this
test, the algorithm fails right away.  Otherwise we go on to the second
test, denoted by {\em matching test} M-T{\footnotesize EST}$(G_i,I)$, which
tries to find a valid spanning star forest.  If this test
succeeds, we are done; otherwise the failure of this test yields a way to
augment $I$ and we proceed to the next iteration.  The algorithm is
outlined in Algorithm~\ref{alg:lMCenter}.

\begin{figure}[t]
    \centerline{\includegraphics[width=0.9\linewidth]{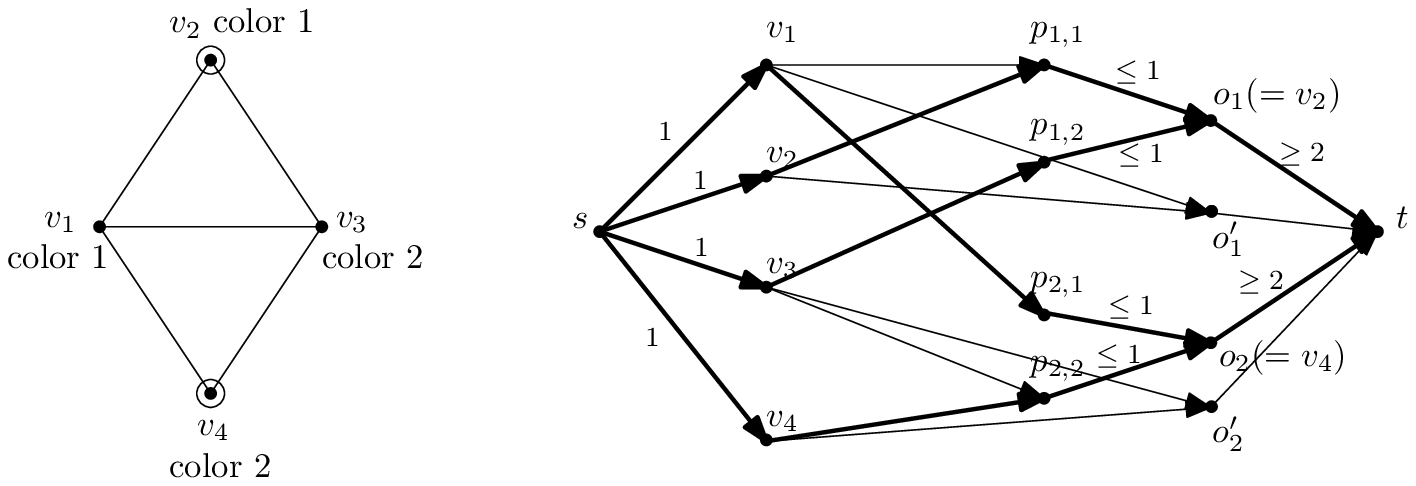}}
    \caption{The flow network construction. On the left is the original
      graph, $I = \{v_2, v_4\}$, $\ell = 2$. On the right is the
      corresponding flow network. Thick edges denote a feasible flow of
      value $|I|\ell = 4$.}
    \label{fig:flow}
\end{figure}

\linesnumbered
\begin{algorithm}[t]
\caption{Algorithm to find an $I$ in $G_i$ to meet condition (II)}
\label{alg:lMCenter}
Let $I$ be an arbitrary maximal independent set in $G_i$\;
\While { {\rm \flowtest($G_i, I$) is passed}} {
  $(S,S')\leftarrow$ \matchtest($G_i$,$I$)\,\, /* $S\subset V,S'\subseteq I$ */\;
  \eIf{$S =\emptyset$}{
    Succeed;
  }{
    $I\leftarrow I-S'+S$\;
    Add nodes to $I$ until it is a maximal independent set\;
  }
}
Fail;\
\end{algorithm}
We now elaborate on \flowtest\ and \matchtest.  \flowtest($G_{i}, I$) checks
if there is a spanning star forest in $G_{i}$ with $I$ being the star centers
such that each star contains at least $\ell$ colors.  As the name
suggests, we conduct the test using a network flow computation.  We first
create a source $s$ and a sink $t$.  For each node $v\in V$, we
add an edge $(s,v)$ with capacity $1$, and for each node $o_j\in I
(1 \le j \le |I|)$, we create a vertex $o_j$ and add an outgoing
edge $(o_j,t)$ with capacity lower bound $\ell$. For each node
$o_j\in I$ and each color $c$, we create a vertex $p_{j,c}$ and an
edge $(p_{j,c},o_j)$ with capacity upper bound $1$. For any $v \in
V$ such that $(v,o_j)\in E_i$ or $v = o_j$, and $v$ has color
$c$, we add an edge from $v$ to $p_{j,c}$ without capacity
constraint. Finally, we add one new vertex $o'_j$ for each $o_j\in
I$, connect all $v \in V$ to $o'_j$ without capacity constraint if
$(v,o_j)\in E$ or $v = o_j$, and connect $o'_j$ to $t$ without
capacity constraint.  
The capacity upper bound of $(p_{j,c}, o_{j})$ forces
at most one node with color $c$ to be assigned to $o_{j}$. 
Therefore, all nodes assigned to $o_{j}$ have distinct colors.
The capacity lower bounds of $(o_{j}, t)$s require that each cluster has
at least $\ell$ nodes. Nodes $o'_{j}$s are used to absorb
other unassigned nodes.
It is not difficult to see that there exists
a semi-valid spanning star forest with nodes in $I$ being star
centers in $G_i$ if an $n$-units flow can be found. In this case
we say that the \flowtest\ is passed.  
See Figure~\ref{fig:flow}
for an example.  Note that a network flow problem with both
capacity lower bounds and upper bounds is usually referred to as the
{\em circulation problem}, and is polynomially solvable~\cite{Korte:07}.

Once $G_i$ and $I$ pass \flowtest, we try to redistribute those
vertices that cause color conflicts.  We do so by a bipartite
matching test \matchtest($G_{i}, I$) which returns two vertex sets
$S$ and $S'$ that are useful later.
Concretely, we test whether there
exists a matching in the bipartite graph
$B(I-C,C-I;E_{G_i}(I-C;C-I))$ for each color class $C$ such that
all vertices in $C-I$ are matched. If such matchings can be found
for all the colors, we say that the \matchtest\ is passed.  Note
that all these matchings together give a spanning star forest such
that each star is polychromatic.  However, this does not guarantee
that the cardinality constraint is preserved.  The crucial fact
here is that $I$ passes {\em
  both} \flowtest\ and \matchtest.  In Lemma \ref{lem:test} we formally
prove that there exists a valid spanning star forest with nodes in $I$ as
star centers if and only if $G_i$ and $I$ pass both \flowtest\ and
\matchtest .  To actually find a valid spanning star forest, we can again
use the network flow construction in \flowtest\ but without the $o'_j$
nodes.  If \matchtest\ fails, we know that for some color class $C$, there
exists a subset $S\subseteq C-I$ such that the size of its neighbor set
$|N_B(S)|$ is less than $|S|$ by Hall's theorem \cite{Korte:07}.  In this
case, \matchtest\ returns $(S,N_B(S))$; such a set $S$ can be found by a
maximum matching algorithm.  Then we update the independent set
$I\leftarrow I-N_B(S)+S$; we show that $I$ is still an independent set in
Lemma~\ref{lem:IS}.  Finally, we add nodes to $I$ arbitrarily 
until it becomes a maximal independent set.
Then, we start the next iteration with the new $I$.  Since $|S| >
|N_B(S)|$, we increase $|I|$ by at least one in each iteration. So the
algorithm terminates in $\leq n$ iterations.

%We will prove in the main theorem that $G_{i^*}$ and some

Before proving that Algorithm~\ref{alg:lMCenter} is guaranteed to succeed
when $w(e_i) = d^*$, we prove the two lemmas left in the description of
our algorithm. The first lemma ensures that $I$ is always an independent
set.

\begin{lemma}
\label{lem:IS} The new set $I \leftarrow I-S'+S$ obtained in each update is
still an independent set in $G_i$.
\end{lemma}
\begin{proof}
Since all vertices in $S$ have the same color, there is no edge
among them. Therefore, we only need to prove that there is no edge
between $I-S'$ and $S$, which is trivial since $S'=N_B(S)$.\qed
\end{proof}

\noindent The second lemma guarantees that we find a feasible solution
if both tests are passed.
\begin{lemma}
\label{lem:test} Given $G_i = G(V,E_i)$ and $I$, a maximal
independent set of $G_i$,  both \flowtest($G_i$,$I$) and
\matchtest($G_i$,$I$) are passed if and only if there exists a valid spanning
star forest in $G_i$ with nodes in $I$ being star centers.
\end{lemma}
\begin{proof}
  The ``if'' part is trivial. We only prove the ``only if'' part.  Suppose
  $G_i$ and $I$ pass both \flowtest($G_i$,$I$) and \matchtest($G_i$,$I$).
  Consider a semi-valid spanning star forest obtained in $G_i$ after
  \flowtest. We delete a minimal set of leaves to make it a valid star (not
  necessarily spanning) forest $F$. Consider the bipartite graph
  $B(I-C;C-I,E_{G_i}(I-C;C-I))$ for each color class $C$. We can see $F\cap
  B$ is a matching in $B$. Since $I$ passes \matchtest, we know that there
  exists a maximum matching such that all nodes in $C-I$ can be matched. If
  we use the Hungarian algorithm to compute a maximum matching with $F\cap
  B$ as the initial matching, the nodes in $I-C$ which are originally
  matched will still be matched in the maximum matching due to the property
  of the alternating path augmentation\footnote{Recall that an augmenting
    path $P$ (with respect to matching $M$) is a path starting from
    unmatched node, alternating between unmatched and matched edges and
    ending also at an unmatched node (for example, see
    \cite{book/algorithm}).  By taking the symmetric difference of $P$ and
    $M$, which we call augmenting on $P$, we can obtain a new matching with
    one more edge.  }.  Therefore, the following invariants are maintained:
  each star is polychromatic and has at least $\ell$ colors. By applying
  the above maximum matching computation for each color class, we obtain a
  valid spanning star forest in $G_i$.  \qed
\end{proof}

Finally, we prove that Algorithm~\ref{alg:lMCenter} is guaranteed to
succeed on $G_{i*}$ for the maximal index $i^*$ such that $w(e_{i^*}) =
d^*$, where $d^*$ is the optimal cluster diameter of any valid spanning
star forest of $G$.

\begin{lemma}
\label{lem:succeed}
Algorithm~\ref{alg:lMCenter} will succeed on $G_{i^*}$.
\end{lemma}
\begin{proof}
  Suppose $\calC^*=\{C^*_1,\ldots,C^*_{k^*}\}$ is the set of clusters
  in the optimal clustering with cluster diameter $d^*$.  Since $G_{i*}$
  include all edges of weights no more than $d^*$, each $C^*_j$ induces a
  clique in $G_{i^*}$ for all $1\leq j\leq k^*$, thus it contains at most
  one node in any independent set. Therefore, any maximal independent set
  $I$ in $G_{i^*}$ can pass \flowtest($G_{i^*},I$), and we only need to
  argue that $I$ will also pass \matchtest($G_{i^*},I$).  Each update to
  the independent set $I$ increases the size of $I$ by at least $1$ and the
  maximum size of $I$ is $k^*$.  When $|I|=k^*$, each $C^*_j$ contains
  exactly one node in $I$ and this $I$ must be able to pass
  \matchtest($G_{i^*},I$).  So Algorithm~\ref{alg:lMCenter} must succeed in
  some iteration.~\qed
\end{proof}

By Lemma~\ref{lem:succeed}, the cost of the spanning star forest found by
Algorithm~\ref{alg:lMCenter} is at most $d^*$.  Since the cost of the
optimal spanning forest is at least $d^*/2$, we obtain a 2-approximation.

\begin{theorem}
\label{thm:2approx}
There is a polynomial-time 2-approximation for \lMCenter.
\end{theorem}

%The next theorem could be seen as a direct corollary of
%theorem~\ref{thm:2approx}.

%\begin{theorem}
%There is an $O(\ell)$-approximation algorithm for the \lTCenter\
%problem.
%\end{theorem}

%%% Local Variables:
%%% mode: latex
%%% TeX-master: "paper"
%%% End:

%% file: lowerBound.tex
\section{The Lower Bound}
\label{sec:lowerbound}
In this section, we show that \lMCenter\ is NP-hard to approximate
within a factor less than $2$ even when there are only three colors.
Therefore the approximation ratio given in the Section~\ref{sec:algorithm} for \lMCenter\ is tight.
Note that if there are two colors, the problem is polynomially
solvable. Indeed, if there are only two colors
%, $\ell$ must be $1$ or $2$. The former case is meaningless. For the later case,
we can use the the following simple algorithm to obtain an optimal
solution.
We start with an empty graph and add edges one by one in an increasing order
of their weights (as before, we just add those edges
whose endpoints' colors are different), getting a series of threshold
graphs $G_i (1 \le i \le m)$.
For each graph $G_i$, we try to find a perfect matching between the
two color classes.  It is easy to see $w(e_i)$ is the optimal
solution where $i$ is the smallest such that a perfect matching in
$G_i$ exists.

\begin{theorem}
\label{thm:lowerbound}
There is no polynomial-time approximation algorithm for
\lMCenter\ that achieves an approximation factor
less than $2$ unless $P=NP$.
\end{theorem}

To prove the NP-hardness for three colors, we show first the
following problem is NP-hard: Given a $3$-colorable graph $G =
(V,E)$ and a feasible $3$-coloring, decide if $V$ can be partitioned
into subsets of size three such that the three vertices in each
subset are connected and the colors of them are all different.  We
call such a partition a {\em
$P(ath)_3$-partition} of $G$.  Note that the NP-hardness of the
$P_3$-partitioning problem directly leads to the fact that \lMCenter\
cannot be approximated within a factor less than 2, since if we assign the
weights of all edges in $G$ to be 1 and consider the metric
completion\footnote{The metric completion of $G(V,E)$ is a complete graph
  with vertex set $V$ and the weight of edge $(u,v)$ defined by the
  shortest path distance between $u$ and $v$ in $G$ for every $u,v\in V$.}
of $G$, then the optimal solution of \lMCenter\ on $G$ is $1$ if $G$ admits
a $P_3$-partitioning and at least $2$ otherwise.

We reduce from the well-known {\em $3$-dimensional matching} problem \cite{GareyJohnson79} to
the $P_3$-partition problem.  Recall that in a $3$-dimensional matching instance, we are given
a tripartite hyper-graph $G = (X \cup Y \cup Z,E)$ with color classes $X,Y,Z$ such that
$|X|=|Y|=|Z|$. Each hyper-edge is of the form
$(x,y,z), x\in X, y\in Y, z\in Z$.  A perfect matching is a set $M
\subseteq E$ of hyper-edges such that each vertex is incident to exactly one
edge in $M$. Given a $3$-dimensional matching instance $G(V,E)$, we
construct a $3$-colorable graph $G'$ as well as a feasible $3$-coloring
such that $G$ has a perfect matching if and only if $G'$ can be
$P_3$-partitioned.

\begin{figure}[t]
    \centerline{\includegraphics[width=\linewidth]{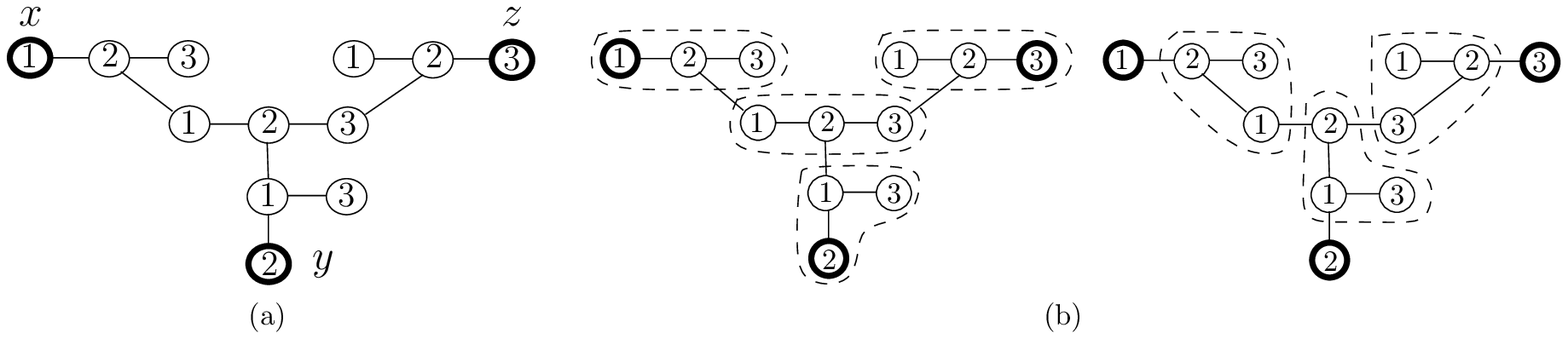}}
    \caption{(a)the gadget; (b) two possible partitions of the gadget. Nodes in thick circle are {\em corner nodes}.}
    \label{fig:gadget}
\end{figure}

The key component of our reduction is the gadget depicted in
Figure \ref{fig:gadget}.
$G'$ has a copy
of the vertices of $G$.
We color the vertices in $X, Y$ and $Z$ with color $1$,$2$ and $3$, respectively.
%The number associated with each node
%denotes the color we have assigned to that node.
For each hyperedge $e=(x, y, z)\in E(G)$, we
attach a distinct gadget to $G'$ by identifying $x, y, z$ with the three corner nodes of the gadget, respectively
(corner nodes are those in thick circles in Figure \ref{fig:gadget}(a)).
The gadget has the following nice property.

\begin{property}
  If $G'$ can be $P_3$-partitioned, then any $P_3$-partition of $G'$
  restricted in one gadget can only take one of the two forms shown in
  Figure \ref{fig:gadget}(b).
\end{property}

The proof of the property is graphically obvious owing to the structure
of the gadget. One can easily make it rigorous by a case by case analysis.
To relate the partition of each gadget to the 3-dimensional matching
problem, we use the following reduction. We take
$e(x, y ,z)$ as a matching edge if and only if the $P_3$-partition on $G'$ restricted
in the corresponding gadget takes the
first form in Figure \ref{fig:gadget}(b).
%, otherwise we don't take the matching.
It is easy to see that $G$ has a perfect matching if
and only if $G'$ can be $P_3$-partitioned.

%%% Local Variables:
%%% mode: latex
%%% TeX-master: "paper"
%%% End:

%% file: outlier.tex
\section{Dealing with Unqualified Inputs}
\label{sec:outliers}
For the \lMCenter\ problem, a feasible solution may not
exist depending on the input color distribution.  The following simple lemma
gives a necessary and sufficient condition for the existence of a feasible
solution. 
%Due to space constraint, all missing proofs are provided in the appendix.

\begin{lemma}
\label{lem:necsuf} There exists a feasible solution for \lMCenter\
if and only if the number of nodes with the same color $c$ is at
most $\lfloor {n\over \ell} \rfloor$ for each color $c$.
\end{lemma}

\begin{proof}
The ``only if'' part is trivial. We only show the ``if'' part.
Suppose the color class $C$ contains the most number of nodes. We
create $|C|$ empty clusters, and process color classes one after
another. For each node $v$, we put it into the cluster currently
containing the least number of nodes provided that the cluster does
not contain a node having the same color as $v$. Note that during
the process, it can be easily shown that the sizes of clusters
differ by at most one by induction. Therefore, each cluster contains
at least $\left\lfloor{n\over |C|}\right\rfloor\geq \ell$ nodes at
the end of the process. \qed
\end{proof}

\noindent To cluster an instance without a feasible solution,
we must exclude some nodes as {\em outliers}. The following lemma
characterizes the minimum number of outliers.% we must exclude.

\begin{lemma}
\label{lem:minoutlier}
Let $C_1,C_2,\ldots,C_k$ be the color
classes sorted in the non-increasing order of their sizes.
\begin{enumerate}
\item Let $p$ be the maximum integer satisfying $\sum_{i=1}^k \min\left( p,
|C_i|\right)\geq p\ell$. The minimum number of outliers is given by
$q=\sum_{i=1}^k \max\left( 0, |C_i|-p\right)$ and $p$ is the
number of clusters when we exclude $q$ outliers.
\item $p\ell\leq n-q <p(\ell+1)$.
\end{enumerate}
\end{lemma}

\begin{proof}
  Let $p^*$ be the maximum number of clusters in any feasible solution.  It
  is easy to see that we must exclude at least $\sum_{i=1}^k \max\left( 0,
    |C_i|-p^*\right)$ outliers. Thus the number of points left is at most
  $$\sum_{i=1}^k \left(|C_i|-\max\left(0,
      |C_i|-p^*\right)\right)=\sum_{i=1}^k \min\left( p^*, |C_i|\right),$$
  which should be at least $p^*\ell$.  Since we pick $p$ to be the maximum
  integer such that $\sum_{i=1}^k \min\left( p, |C_i|\right)\geq p\ell$, we
  have $p\geq p^*$.  Thus $q$ is a lower bound on the number of outliers.
  Moreover, if we delete $\max\left( 0, |C_i|-p\right)$ nodes from $C_i$,
  the remainders admit a feasible clustering from Lemma \ref{lem:necsuf}.
  Therefore, $q$ is also a upper bound, and the proof of part 1 is completed.

  $n-q\geq p\ell$ is obvious. The $n-q<p(\ell+1)$ can be seen as
  follows. Suppose it is not true. We have
  $$
    n-\sum_{i=1}^k \max (0,|C_i|-p) = \sum_{i=1}^k (|C_i|-\max(0,|C_i|-p))
    =\sum_{i=1}^k \min(p,|C_i|)\geq (p+1)\ell.
  $$
  Thus, $\sum_{i=1}^k \min(p+1,|C_i|)\geq (p+1)\ell$ which
  contradicts the maximality of $p$.
  \qed
\end{proof}

With lemma~\ref{lem:minoutlier} at hand, it is natural to consider
the following optimization problem: find an \lMCenter\ solution by
clustering $n - q$ points such that the maximum cluster radius is
minimized. We call this problem \lMCenterOutlier.
From Lemma \ref{lem:minoutlier} we can see that $p$ is
independent on the metric and can be computed in advance.  In
addition, implicit from Lemma~\ref{lem:minoutlier} is that the
number of outliers of each color is also fixed, but we need to
decide {\em which} points should be chosen as outliers.

In the fortunate case where we have a color class $C$ with exactly $p$
nodes, we know that there is exactly one node of $C$ in each cluster
of any feasible solution.  By using a similar flow network construction
used in \flowtest, we can easily get a $2$-approximation using $C$ as
the cluster centers.  However, the problem becomes much more difficult when
the sizes of all color classes are different from $p$.  Loosely
speaking, our difficulty is two-fold: exponentially many choices of
outliers and cluster centers.
%and exponentially many choices of cluster centers.
%If any one of them can be fixed, we can easily devise a constant approximation by using
%the techniques developed in Section \ref{sec:algorithm}.
\subsection{A Constant Approximation}
We first define some notations.
We call color classes of size larger than $p$ {\em popular colors}
and nodes having such colors {\em popular} nodes. Other color classes have
at most $p$ nodes, and these nodes are {\em unpopular}.  We denote
the set of popular nodes by $\calP$ and the set of unpopular nodes
by $\calN$. Note that after removing the outliers, each popular color
has exactly $p$ nodes and each cluster will contain the same number
of popular nodes. Let $z$ be the number of popular nodes each
cluster contains. We denote by $G^d$ the power graph of $G$ in which
two vertices $u,v$ are adjacent if
there is path connecting $u$ and $v$ with at most $d$ edges.
The length of the edge $(u,v)$ in $G^d$ is set to be $\dist_G(u,v)$.
Before describing the algorithm, we need the following simple lemma.

\begin{lemma}
\label{lm:hamiltonian} For any connected graph $G$, $G^3$ contains a
Hamiltonian cycle which can be found in linear time.
\end{lemma}

\begin{proof}
Let $T$ be any spanning tree of $G$. It suffices to prove $T^3$
contains a Hamiltonian cycle. We root $T$ at an arbitrary vertex $r$
and denote the depth of a vertex $v$ by $\depth(v)$ ($\depth(r)=1$).
Let $T_v$ be the subtree rooted at $v$ and $Ch(v)$ be the set of
children of $v$. Consider the Algorithm \ref{alg:traverse}:
{\em Traverse}$(r)$. Clearly, the algorithm runs in linear time.

Suppose the order obtained is $\{v_1,v_2,\ldots,v_n\}$. We claim
$\dist_T(v_i,v_{i+1})\leq 3$ for $1\leq i\leq n-1$ and
$\dist_T(v_1,v_n)\leq 1$. Note that the claim immediately implies
the lemma. We prove the claim by induction on the size of the tree.
Suppose $\depth(v)$ is odd (the other case can be proved similarly)
and $Ch(v)=\{u_1,\ldots,u_k\}$. Let
$O_i=\{v_{i,1},v_{i,2},\ldots,v_{i,|O_i|}\}$ be the traverse order
of $T_{u_i}$. It is easy to see the order produced by
{\em Traverse}$(v)$ is $v, O_1, O_2,\ldots, O_k$.
%By induction hypothesis,we only need to prove the claim when
It is also not hard to see that 
$$\dist_T(v,v_{1,1})\leq
\dist_T(v,u_1)+\dist_T(u_1,v_{1,1})
=\dist_T(v,u_1)+\dist_T(v_{1,|O_1|},v_{1,1})\leq 2$$
$$\text{and}\;\;\;\;\dist_T(v_{k,|O_k|},v)=\dist_T(u_k,v)=1$$ since $u_i$ is last vertex
in order $O_i$. We can also see that
 $$\dist_T(v_{i,|O_i|},v_{i+1,1})\leq
\dist_T(u_i,v)+\dist_T(v,u_{i+1})+\dist_T(u_{i+1},v_{i+1,1})\leq 3.$$
By induction hypothesis, the proof is completed. \qed
\end{proof}

\linesnumbered
\begin{algorithm}[t]
\caption{Traverse($v$)} \label{alg:traverse}
  \eIf{ $\depth(v)$ is odd} {
    visit($v$)\;
    \textbf{for each} ($u\in Ch(v)$) Traverse($u$)\;
  }{
    \textbf{for each} ($u\in Ch(v)$) Traverse($u$)\;
    visit($v$)\;
  }
\end{algorithm}

The algorithm still adopts the thresholding method, that is, we add
edges one by one to get graphs $G_i =
(V,E_i=\{e_1,e_2,\ldots,e_i\})$, for $i =1,2,\dots$, and in each
$G_i$, we try to find a valid star forest that spans $G_i$ except
$q$ outliers.  Let $d^*$ be the diameter of the optimal solution
that clusters $n-q$ points, and $i^*$ be the maximum index such
that $w(e_i) = d^*$.
Let $G_i[\calN]$ be the subgraph of $G_i$ induced by all unpopular nodes.
We define the ball of radius $r$ around $v$
to be $\B(v,r)=\{u\in \calN \mid \dist_G(v,u)\leq r\}$.
For each $G_i$, we run the Algorithm: \lMCenterOutlier$(G_i)$ (see below).
We proceed to $G_{i+1}$ when the algorithm claims failure.

The high level idea of the algorithm is as follows:
Our goal is to show that the algorithm can find a valid star forest spanning $n-q$
nodes in $G_{i^*}^{28}$.
It is not hard to see that this gives us an approximation algorithm
with factor $28\times 2=56$.
First, we notice that \flowtest\ can be easily modified to work for the outlier version
by excluding all $o'_j$ nodes and testing whether there is a flow of value $n-q$.
However, the network flow construction needs to know in advance the set of candidates of cluster centers.
For this purpose, we attempt to attach a set $U$ of
$p$ new nodes which we call {\em virtual centers} to $G_i$ which serve as the candidates of cluster centers in \flowtest.
In the ideal case, if these virtual centers can be distributed such that each of them is attached to a distinct optimal cluster,
\flowtest\ can easily produce a 2-approximation.
Since the optimal clustering is not known, this is very difficult in general.
However, we show that there is way to carefully distribute the virtual centers
such that there is a perfect matching between these virtual centers and the optimal cluster centers
and the longest matching edge is at most $27d^*$.
This implies that there is a valid spanning star forest
in $G^{27}_{i^*}$ (each star is formed by an optimal cluster
together with the virtual center that matches the cluster center).
Also, it is easy to see that each virtual center is at most $27d^{*}+d^{*}$ away from any other node
in the same star.
Therefore, it suffices to just run \flowtest($G^{28}_{i^{*}},U$) 
to find a feasible solution.

\paragraph{{\bf Algorithm}: \lMCenterOutlier$(G_i)$.}
\begin{enumerate}
\item If $G_i[\calN]$ contains a connected component with less
than $\ell-z$ nodes, we declare failure.

\item Pick an arbitrary unpopular node $v$ such that $|\B(v, w(e_i))|\geq
  \ell-z$ and delete all vertices in this ball; repeat until no such node
  exists.  Then, pick an arbitrary unpopular node $v$ and delete all
  vertices in $\B(v, w(e_i))$; repeat until no unpopular node is left. Let
  $\B_1,\B_2,\ldots, \B_k$ be the balls created during the process. If a
  ball contains at least $\ell-z$ unpopular nodes, we call it {\em big}.
  Otherwise, we call it {\em small}.

\item In $G_i[\calN]$, shrink each $\B_j$ into a single node
  $b_j$. A node $b_j$ is {\em big} if $\B_j$ is big and
  {\em small} otherwise. We define the {\em weight} of $b_j$ to be
  $\mu(b_j)={|\B_j|\over \ell-z}$. Let the resulting graph with vertex set
  $\{b_j\}_{j=1}^k$ be $D_i$.

\item For each connected component $C$ of $D_i$, do

\begin{enumerate}
    \item Find a spanning tree $T_C$ of $C^3$ such that all small nodes are
    leaves. If this is not possible, we declare failure.
   
    \item Find (by Lemma \ref{lm:hamiltonian}) a Hamiltonian cycle
    $P=\{b_{1},b_{2},\ldots, b_{h},b_{h+1}=b_{1}\}$ over all non-leaf nodes
    of $C$ such that $\dist_{D_i}(b_{j},b_{j+1})\leq 9w(e_i)$.
\end{enumerate}

\item We create a new color
  class $U$ of $p$ nodes which will serve as ``virtual centers'' of
  the $p$ clusters.
  %Correspondingly, we will require each cluster to contain at
  %least $\ell$ nodes of distinct colors excluding its virtual center.
  These virtual centers are placed in $G_i$ ``evenly" as follows.  Consider
  each connected component $C$ in $D_i$ and the corresponding spanning tree
  $T_C$ of $C^3$.  For each non-leaf node $b_j$ in $T_C$, let $L(b_j)$ be
  the set of leaves connected to $b_j$ in $T_C$, and let
  $\eta(b_j)=\mu(b_j)+\sum_{b_x\in L(b_j)}\mu(b_x)$ and
  $\delta_j=\sum_{x=1}^{j} \eta(b_x)$.  We attach $\lfloor
  \delta_{i}\rfloor-\lfloor \delta_{i-1}\rfloor$ virtual centers to the
  center of $\B_i$ by zero weight edges.  If the total number of virtual
  centers used is not equal to $p$, we declare failure. Let $H_i$ be the
  resulting graph (including all popular nodes, unpopular nodes and virtual
  centers).

\item Find a valid star forest in $H_i^{28}$ using $U$ as centers, which spans
  $n-q$ nodes (not including the nodes in $ U$) by using \flowtest.
  If succeeds, we return the star forest found, otherwise we declare failure.
\end{enumerate}

\subsection{Analysis of the algorithm}
We show that the algorithm succeeds on $G_{i^*}$.
%thus the
%approximation ratio would be 2 (by theorem~\ref{thm:2approx}) times 8
Since we perform \flowtest\ on $H_{i^*}^{28}$ in which each edge is of
length $\leq 28d^*$, the radius of each cluster is at most $28d^*\leq 56r^{*}$.
Therefore, the approximation ratio is $56$.

Let $H_{i^*}$ be the graph obtained by adding virtual centers to
$G_{i^*}$ as described above.
Let $\calC^*=\{C_1^*, \ldots,C_p^*\}$ be the optimal clustering.
Let $I^*=\{\nu^*_1,\ldots, \nu^*_p\}$ be the set of cluster centers of $\calC^*$ where
$\nu^*_i$ is the center of $C_i^*$.  We denote the balls grown in step 2
by $\B_1,\ldots, \B_k$. Let $\nu_i$ be the center of $\B_i$.

The algorithm may possibly fail in step 1, step 4(a), step 5 and step 6.
Obviously $G_{i^*}$ can pass step 1.
Therefore, we only check the other three cases.

\topic{Step 4(a)} We prove that the subgraph induced by all big nodes are
connected in $C^3$.  Indeed, we claim that each small node is adjacent to
at least one big node in $C$ from which the proof follows easily.  Now we
prove the claim.  Suppose $b_j$ is a small node and all its neighbors are
small.  We know that in $G_{i^*}[\calN]$, $\nu_j$ has at least $\ell-z-1$
neighbors because $\nu_j$ is an unpopular node and thus belongs to some
optimal cluster. So we could form a big ball around $\nu_j$, thus
contradicting to the fact that $\nu_j$ is in a small ball.  To find a
spanning tree with all small nodes as leaves, we first assign each small
node to one of its adjacent node arbitrarily and then compute a tree
spanning all the big nodes.

\topic{Step 5}
We can see that in each connected component $C$ (with big nodes $b_1,\ldots,b_h$) in $D_{i^*}$,
the total number of virtual centers we have placed is
$
\sum_{i=1}^h(\lfloor \delta_{i}\rfloor-\lfloor \delta_{i-1}\rfloor)=\lfloor \delta_h\rfloor=\lfloor \sum_{x=1}^h \eta(b_x)\rfloor
=\lfloor \sum_{b_j\in C} \mu(b_j) \rfloor =\left\lfloor  \frac{|C|}{l-z } \right\rfloor
$
where $|C|=\sum_{b_j\in C}|\B_j|$, the number of nodes in the connected component
of $G_{i^*}[\calN]$ corresponding to $C$.
This is at least the number of clusters
created for the component $C$ in the optimal solution.
Therefore, we can see the total number of virtual
centers created is at least $p$. On the other hand, from
Lemma~\ref{lem:minoutlier}(2), we can see that $p(\ell-z)\leq |\calN|<(p+1)(\ell-z)$.
Hence,
$
p=\left\lfloor{|\calN|\over \ell-z}\right\rfloor
=\left\lfloor{\sum_{C}|C|\over \ell-z}\right\rfloor
    \geq
    \sum_{\textrm{C}} \left\lfloor {|C|\over \ell-z}\right\rfloor.
$
where the summation is over all connect components. So, we prove that exactly $p$ virtual centers
were placed in $G_{i^*}$.

\topic{Step 6} According to the high level idea discussed before, we only
need to show that there is a perfect matching $M$ between $U$ and the set
of optimal centers $I^*$ in $H^{27}_{i^*}$.  
%% For each $\nu^*_i\in I^*$, we
%% denote its matching node in $U$ by $M(\nu^*_i)$.  The success of the
%% algorithm follows since there exists a star forest in which each star
%% consists of one $M(\nu^*_i)\in U$ and nodes in a distinct optimal cluster
%% $C_i^* \in \calC^*$.  It is easy to see $M(\nu^*_i)$ is at most
%% $27d^{*}+d^{*}=28d^*$ away from each node in $C_i^*$.  To show the
%% existence of the perfect matching $M$, 
We consider the bipartite subgraph $Q( U,I^*, E_{H^{27}_{i^*}}( U, I^*))$.
From Hall's theorem, it suffices to show that $|N_Q(S)|\geq |S|$ for any
$S\subseteq U$, which can be implied by the following lemma.

\begin{lemma}
\label{lm:nei}
For any $S\subseteq U$, the union of the balls of radius $27d^*$ around
the nodes of $S$, i.e, $\bigcup_{u\in S}\B(u,27d^*)$, intersects at
least $|S|$ optimal clusters in $\calC^*$.
\end{lemma}
\begin{proof}
  We can assume w.l.o.g. all nodes of $S$ are in a single connected component of $G_{i^*}[\calN]$.
  The generalization to several connected components is straightforward.
  Let $P=\{b_1, b_2 \ldots, b_h\}$ be the Hamiltonian cycle (found in Step 4(b)) for such a component
  (actually, the component after shrinking balls).
  Let $\tilde P$ be the
  set of nodes $b_j\in P$ such that at least one virtual center in $S$ is attached to $\B_j$.

  We first assume $|\tilde P|\leq h-2$.
  In this case, we claim that $\left|\bigcup_{u \in S} \B(u , 27d^*)\right| > (|S|+1)(\ell-z).$
  We know by definition that the number of
  nodes in each big ball $\B_j$ plus nodes in those small balls $\B_x$
  attached to it (that is, $b_x \in L(b_j)$) is $\eta(b_j)(\ell-z)$.
  $\tilde P$ can be seen as a collection of subpaths of $P$. For each of those subpaths,
  say $\{b_j,\ldots,b_{j'}\}$, the number of nodes in $S$ attached to it is at most
  $$
  \sum_{x=j}^{j'}(\lfloor \delta_{x}\rfloor-\lfloor \delta_{x-1}\rfloor)=
  \lfloor  \delta_{j'}\rfloor-\lfloor \delta_{j-1}\rfloor=
  \lfloor  \sum_{x=1}^{j'}\eta(b^*_x)\rfloor-\lfloor \sum_{x=1}^{j}\eta(b^*_x)\rfloor
  \leq \sum_{x=j}^{j'}\eta(b^*_x)+1
  $$
  On the other hand, we can see that $\B(\nu_j,27d^*)$ contains
  all nodes in $\B_{j-1}$,$\B_j$ and $\B_{j+1}$ \footnote{ $\B_0=\B_h,
  \B_{h+1}=\B_1$.  }  and all nodes in the small balls attached to $\B_j$.
  This is because $\dist_{D_{i^*}}(b_{j},b_{j+1})\leq 9d^*$
  and each $b_j$ is obtained by shrinking a ball of radius at most $d^*$.
  Therefore,
  $$
  \left|\bigcup_{x=j}^{j'} \B(\nu_x, 27d^*)\right|\geq
  \sum_{x=j-1}^{j'+1}\eta(b^*_x) (\ell-z) \geq (\sum_{x=j}^{j'}\eta(b^*_x)
  +2)(\ell-z) > (\lfloor \delta_{j'}\rfloor-\lfloor \delta_{j-1}+1\rfloor)(l-z)
  $$
  where the second inequality holds since $\eta(b_j)\geq 1$ for any big node $b_j$.
  Summing up all the subpaths of $\tilde P$ proves the claim.
  From Lemma \ref{lem:minoutlier}.2 and the fact that each cluster
  has exactly $z$ popular nodes, we can see $|S|$ optimal clusters contains less than $(\ell-z)(|S|+1)$ unpopular nodes.
  Therefore, the lemma holds.

  If $|\tilde P|>h-2$, then $\bigcup_{u\in S} \B(u,27d^*)$ contains all
  unpopular nodes in this component. The lemma also follows. \qed
\end{proof}

%Therefore, the algorithm succeeds on finding a valid star forest which spans $n-q$ nodes on $G^{28}_{i^*}$.
% the maximum diameter of the clustering we found is at most $2\times 28d^*$.

\begin{theorem}
  There is a polynomial-time algorithm for \lMCenterOutlier\ that
  produces a $56$-approximation.
\end{theorem}

\section{Further Directions}
This work results in several open questions.
First, as in \cite{AFK+06}, we could also try to minimize the sum of the radii of
the clusters.  However, this seems to be much more difficult, and we leave
it as an interesting open problem.
Another open problem is to design constant approximations
for the problem with any fixed number of outliers, that is,
for a given number $k$, find an optimal clustering if at most $k$ outliers
can be removed.

As mentioned in the introduction, our work can be seen as a stab at the more 
general problem of clustering under instance-level hard constraints.
Although arbitrary CL (cannot-link) constraints seems hard to approximate with respect to minimizing the number of clusters due to the hardness of graph coloring \cite{davidson2007iac}, 
other objectives and special classes of constraints, e.g. diversity constraints, may still
admit good approximations. Besides the basic ML and CL constraints, 
we could consider more complex constraints like the rules proposed in the
Dedupalog project~\cite{arasu2009large}. One example of such rules
says that whenever we cluster two points $a$ and $b$ together, 
we must also cluster $c$ and $d$. Much less is known for incorporating these types of
constraints into traditional clustering problems and 
we expect it to be an interesting and rich further direction.

%%% Local Variables: 
%%% mode: latex
%%% TeX-master: "paper"
%%% End: 

%% file: paper.bbl
\begin{thebibliography}{10}

\bibitem{AFK+06}
G.~Aggarwal, T.~Feder, K.~Kenthapadi, S.~Khuller, R.~Panigrahy, D.~Thomas, and
  A.~Zhu.
\newblock Achieving anonymity via clustering.
\newblock In {\em Proc. ACM Symposium on Principles of Database Systems}, pages
  153--162, 2006.

\bibitem{aggarwal05:_anony}
G.~Aggarwal, T.~Feder, K.~Kenthapadi, R.Motwani, R.~Panigrahy, D.~Thomas, and
  A.~Zhu.
\newblock Anonymizing tables.
\newblock In {\em Proc. International Conference on Database Theory}, pages
  246--258, 2005.

\bibitem{journal/jacm/ailon08}
N.~Ailon, M.~Charikar, and A.~Newman.
\newblock Aggregating inconsistent information: Ranking and clustering.
\newblock In {\em Journal of the ACM}, volume 55(5), pages 1--27, 2008.

\bibitem{arasu2009large}
A.~Arasu, C.~R{\'e}, and D.~Suciu.
\newblock {Large-scale deduplication with constraints using Dedupalog}.
\newblock In {\em Proc. IEEE International Conference on Data Engineering},
  pages 952--963, 2009.

\bibitem{bairoch1997spp}
A.~Bairoch and R.~Apweiler.
\newblock {The SWISS-PROT protein sequence data bank and its supplement
  TrEMBL}.
\newblock {\em Nucleic acids research}, 25(1):31, 1997.

\bibitem{bansal2004correlation}
N.~Bansal, A.~Blum, and S.~Chawla.
\newblock {Correlation clustering}.
\newblock {\em Machine Learning}, 56(1):89--113, 2004.

\bibitem{beresford03}
A.~Beresford and F.~Stajano.
\newblock Location privacy in pervasive computing.
\newblock {\em Pervasive Computing, IEEE}, pages 46--55, 2003.

\bibitem{c.-w.wong06}
R.~C.-W.Wong, J.~Li, A.-C. Fu, and K.Wang.
\newblock $(\alpha, k)$-anonymity: an enhanced k-anonymity model for privacy
  preserving data publishing.
\newblock In {\em Proc. ACM SIGKDD International Conference on Knowledge
  Discovery and Data Mining}, pages 754--759, 2006.

\bibitem{charikar2001afl}
M.~Charikar, S.~Khuller, D.~Mount, and G.~Narasimhan.
\newblock {Algorithms for facility location problems with outliers}.
\newblock In {\em Proc. ACM-SIAM Symposium on Discrete Algorithms}, pages
  642--651, 2001.

\bibitem{davidson2007iac}
I.~Davidson and S.~Ravi.
\newblock {Intractability and clustering with constraints}.
\newblock In {\em Proc. International Conference on Machine Learning}, pages
  201--208, 2007.

\bibitem{dwork09:_when}
C.~Dwork, M.~Naor, O.~Reingold, G.~Rothblum, and S.~Vadhan.
\newblock On the complexity of differentially private data release: efficient
  algorithms and hardness results.
\newblock In {\em Proc. ACM Symposium on Theory of Computation}, pages
  381--390, 2009.

\bibitem{feldman09:_privat}
D.~Feldman, A.~Fiat, H.~Kaplan, and K.~Nissim.
\newblock Private coresets.
\newblock In {\em Proc. ACM Symposium on Theory of Computation}, pages
  361--370, 2009.

\bibitem{GareyJohnson79}
M.~R. Garey and D.~S. Johnson.
\newblock Computers and intractability: A guide to the theory of
  np-completenes.
\newblock 1979.

\bibitem{ghinita07:_fast}
G.~Ghinita, P.~Karras, P.~Kalnis, and N.~Mamoulis.
\newblock Fast data anonymization with low information loss.
\newblock In {\em Proc. International Conference on Very Large Databases},
  pages 758--769, 2007.

\bibitem{giotis2006correlation}
I.~Giotis and V.~Guruswami.
\newblock {Correlation clustering with a fixed number of clusters}.
\newblock In {\em Proc. ACM-SIAM Symposium on Discrete Algorithms}, pages
  1176--1185, 2006.

\bibitem{hoppner2008csc}
F.~Hoppner, F.~Klawonn, R.~Platz, and S.~Str.
\newblock {Clustering with Size Constraints}.
\newblock {\em Computational Intelligence Paradigms: Innovative Applications},
  2008.

\bibitem{Ji04}
X.~Ji.
\newblock {\em Graph Partition Problems with Minimum Size Constraints}.
\newblock PhD thesis, Rensselaer Polytechnic Institute, 2004.

\bibitem{kifer06:_injec}
D.~Kifer and J.~Gehrke.
\newblock Injecting utility into anonymized datasets.
\newblock In {\em Proc. ACM SIGMOD International Conference on Management of
  Data}, pages 217--228, 2006.

\bibitem{Korte:07}
B.~Korte and J.~Vygen.
\newblock {\em Combinatorial Optimization: Theory and Algorithms}.
\newblock Springer-Verlag, 4th edition, 2007.

\bibitem{lefevre06:_mondr}
K.~LeFevre, D.~J. DeWitt, and R.~Ramakrishnan.
\newblock Mondrian multidimensional $k$-anonymity.
\newblock In {\em Proc. IEEE International Conference on Data Engineering},
  page~25, 2006.

\bibitem{machanavajjhala06}
A.~Machanavajjhala, J.~Gehrke, D.~Kifer, and M.~Venkitasubramaniam.
\newblock $l$-diversity: Privacy beyond $k$-anonymity.
\newblock In {\em Proc. IEEE International Conference on Data Engineering},
  page~24, 2006.

\bibitem{meyerson04}
A.~Meyerson and R.~Williams.
\newblock On the complexity of optimal $k$-anonymity.
\newblock In {\em Proc. ACM Symposium on Principles of Database Systems}, pages
  223--228, 2004.

\bibitem{book/algorithm}
M.H.Alsuwaiyel.
\newblock {\em Algorithms: Design Techniques and Analysis}.
\newblock World Scienfic, 1998.

\bibitem{HK07}
H.~Park and K.~Shim.
\newblock Approximate algorithms for k-anonymity.
\newblock In {\em Proc. ACM SIGMOD International Conference on Management of
  Data}, 2007.

\bibitem{samarati01:_protec}
P.~Samarati.
\newblock Protecting respondents' identities in microdata release.
\newblock {\em IEEE Transactions on Knowledge and Data Engineering},
  13(6):1010--1027, 2001.

\bibitem{Wagstaff00clusteringwith}
K.~Wagstaff and C.~Cardie.
\newblock Clustering with instance-level constraints.
\newblock In {\em Proc. International Conference on Machine Learning}, pages
  1103--1110, 2000.

\bibitem{wagsta2001ckm}
K.~Wagstaff, C.~Cardie, and S.~Schroedl.
\newblock {Constrained k-means clustering with background knowledge}.
\newblock In {\em Proc. International Conference on Machine Learning}, pages
  577--584, 2001.

\bibitem{xiao06:_anatom}
X.~Xiao and Y.~Tao.
\newblock Anatomy: Simple and effective privacy preservation.
\newblock In {\em Proc. International Conference on Very Large Databases},
  pages 139--150, 2006.

\bibitem{xiao07}
X.~Xiao and Y.~Tao.
\newblock $m$-invariance: Towards privacy preserving re-publication of dynamic
  datasets.
\newblock In {\em Proc. ACM SIGMOD International Conference on Management of
  Data}, pages 689--700, 2007.

\bibitem{Xiao:10}
X.~Xiao, K.~Yi, and Y.~Tao.
\newblock The hardness and approximation algorithms for l-diversity.
\newblock {\em Proc. Conference on Extending Database Technology}, 2010.

\bibitem{xing:dml}
E.~Xing, A.~Ng, M.~Jordan, and S.~Russell.
\newblock {Distance metric learning, with application to clustering with
  side-information}.
\newblock In {\em Proc. Annual Conference on Neural Information Processing
  Systems}, pages 505--512, 2003.

\end{thebibliography}
